\documentclass[preprint,showpacs]{revtex4}
\usepackage{epsfig}
\usepackage{graphicx}
\usepackage{bm}
\newcommand{\bmath}[1]{\mbox{\boldmath{${#1}$}}}

\newcommand{\half}{\mbox{${\textstyle \frac{1}{2}}$}}           

\newcommand{\rd}{\textrm{d}}
\begin{document}
{\flushright{\today}}
\title{Coulomb-nuclear interference in pion-nucleus bremsstrahlung}
\author{G\"oran F\"aldt}\email{goran.faldt@fysast.uu.se} 
\author{Ulla Tengblad}\email{ulla.tengblad@fysast.uu.se} 
\affiliation{ Department of physics and astronomy, 
Uppsala University,
 Box 535, S-751 21 Uppsala,Sweden }

\begin{abstract}
Pion-nucleus bremsstrahlung offers a possibility 
of measuring the structure functions of pion-Compton scattering 
from a study  of the small-momentum-transfer region
where the bremsstrahlung reaction is dominated 
by the single-photon-exchange mechanism. The corresponding 
cross-section distribution
is  characterized by a sharp peak at small momentum transfers.
But there is also a hadronic contribution which is smooth 
and constitutes an undesired background. In this communication 
the modification of the single-photon exchange amplitude 
by multiple-Coulomb scattering is investigated
 as well as the Coulomb-nuclear  interference term. 
\end{abstract}
\maketitle
%
%
%
\section{Introduction}

We shall in this paper add a final touch to the subject of 
hard bremsstrahlung
in pion-nucleus scattering in the Coulomb region. The reaction
studied is 
\[
\pi^- +A\rightarrow\pi^- +\gamma +A  .
\]
In the kinematic region of small momentum transfers to the nucleus
 the reaction is dominated by the 
one-photon-exchange mechanism. We have previously derived 
expressions  \cite{FT1,FT2} both for the {\it Coulomb contribution}, 
i.e.~radiation in conjunction with elastic pion-nucleus-Coulomb scattering ,
and for the {\it nuclear contribution}, i.e.~radiation 
in conjunction with elastic pion-nucleus hadronic scattering. 
Also, detailed predictions for the COMPASS experiment \cite{COMP} at
CERN have been made \cite{FT3}, based on the Coulomb contribution alone.
The aim of the COMPASS experiment  is to investigate the 
electromagnetic structure functions of pion-Compton scattering. 
We studied the sensitivity of pionic bremsstrahlung to details 
of the structure functions
by employing a meson-exchange model for the pion-Compton
amplitudes that in additon 
to the Born contributions contained  contributions from the
 $\sigma$, $\rho$, and $a_1$, exchanges. 

Aspects of the theory that need further investigation
 concern the nuclear background  contribution,
 and the interference between 
Coulomb and nuclear contributions. These aspects are 
investigated in the present paper. Also the
form factor of the Coulomb amplitude, due to multiple 
Coulomb scattering,  is investigated.

This work extends previous theoretical studies by 
Gal'perin et al.\cite{4} and F\"aldt  and Tengblad \cite{FT1,FT2,FT3}.
The results concerning the Coulomb-form factor may 
be important for the interpretation of the data by  
Antipov  et al.\cite{Ant}.

The kinematics of the pion-nucleus bremsstrahlung
 reaction is defined through
\begin{equation}
\pi^-(p_1)+ {\rm A}(p)\rightarrow \pi^-(p_2)+\gamma(q_2)+{\rm} A(p'),
\end{equation}
and the kinematics of the related pion-Compton reaction through
 \begin{equation}
\pi^-(p_1)+ \gamma(q_1)\rightarrow \pi^-(p_2)+\gamma(q_2) , \label{Compt-kin}
\end{equation}
 with $q_1=p-p'$. 
 
Our analysis is carried for high energies and small transverse
momenta, meaning small compared with the longitudinal momenta.
In addition the momentum transfer to the nucleus must be
in the Coulomb region, i.e.~extremely small.

The cross-section distribution is written as 
\begin{equation}
\frac{\rd \sigma}{\rd^2q_{1\bot}  \rd^2q_{2\bot} \rd x}
  = \frac{1}{32 (2 \pi)^5 E_2 \omega_2 M_A^2 }
    \left| \cal{M}_C+  \cal{M}_N\right|^2  ,
 \label{Cross-sect-distr}
 \end{equation}
 where $\cal{M}_C$ is the Coulomb amplitude and 
 $\cal{M}_N$ the nuclear amplitude.
The parameter $x$ is defined as the ratio
\begin{equation}
x=
  \frac{q_{2z}}{p_1} =\frac{\omega_2}{E_1} ,  \label{x-fraction}
\end{equation} so that e.g., $E_2 \omega_2=x(1-x)E_1^2$. 
In hadronic bremsstrahlung there is a  
fixed longitudinal-momentum transfer to the nucleus
which depends on $x$, 
\begin{equation}
q_{1\|}=q_{min}
	  =\frac{m_{\pi}^2}{2E_1}\cdot \frac{x}{1-x}  .\label{qmin-def}
\end{equation}
At high energies  $q_{min}$  is  obviously exceedingly small.

The  structure of the cross-section distribution is mainly determined by
the one-photon-exchange factor
\begin{equation}
\frac{\bmath{q}_{1\bot}^2}{(\bmath{q}_{1\bot}^2+q_{min}^2)^2} \label{Prim_fact}
\end{equation}
which vanishes when the transverse-momentum transfer $\bmath{q}_{1\bot}$
to the nucleus vanishes.  When the momentum transfer to the nucleus, ${q}_{1\bot}$, 
increases far beyond $q_{min}$ we  eventually come to momentum
transfers where
the nuclear  contribution dominates \cite{FT1}. 
%
%
%

\newpage
\section{The Coulomb amplitude}

The expression for the Coulomb amplitude in the one-photon-exchange
approximation is given in Eq.~(23) of ref.\cite{FT2}. It reads
 \begin{eqnarray}
{\cal M}_C^{(B)}&=&\frac{8\pi iZ M_A e\alpha}{\mathbf{q}_1^2} 
   \frac{4xE_2 }{\mathbf{q}_{2\bot}^2 + x^2 m_{\pi}^2 }
 \bigg[ A(x,\mathbf{q}_{2\bot}^2) 
 \bigg(\mathbf{q}_{1\bot} - 2\mathbf{q}_{2\bot}
     \frac{\mathbf{q}_{2\bot}\cdot \mathbf{q}_{1\bot}}
          {\mathbf{q}_{2\bot}^2 + x^2 m_{\pi}^2 }  \bigg)
	      \nonumber \\   &&\nonumber \\
	    &&\quad \quad\qquad\qquad \qquad+ 
	      B(x,\mathbf{q}_{2\bot}^2) \mathbf{q}_{1\bot}
	      \,\bigg]\cdot\bmath{\epsilon}_2 .
	 \label{SimpleII-amp} 
\end{eqnarray} 
The functions $A(x,\mathbf{q}_{2\bot}^2)$ and 
$B(x,\mathbf{q}_{2\bot}^2)$, which were there called 
$\tilde{A}(x,\mathbf{q}_{2\bot}^2)$ and  
$\tilde{B} (x,\mathbf{q}_{2\bot}^2)$,  
are the pion-Compton-structure functions. 
Their analytic expressions  in the 
one-meson-exchange approximation are given in the same reference. In the
Born approximation, i.e.~for point-like pions,
 the structure functions take the values $A=1$ and $B=0$.

As can be inferred from Eq.~(\ref{SimpleII-amp}) the Coulomb 
amplitude can be factorized as
\begin{equation}
	{\cal M}_C^{(B)} = \frac{2Z\alpha}
	   {\bmath{q}_{1\bot}^2+q_{\|}^2}\ 
	   \bmath{g}\cdot\bmath{q}_1, \label{study}
\end{equation}
with the vector $\bmath{g}$  in the impact parameter
plane, so that $ \bmath{g}\cdot\bmath{q}_1= \bmath{g}\cdot\bmath{q}_{1\bot}$. 
This amplitude is valid for a point-like nuclear-charge
distribution. The factor multiplying $\bmath{g}\cdot\bmath{q}_1$
in Eq.(\ref{study}) is the $\pi^{-}$-nucleus-Coulomb-scattering
amplitude in the Born approximation. 

The above expression can be improved by taking into account
the finite extension of the nuclear-charge distribution and
the distortion of the pion trajectory due to Coulomb-multiple
scattering. The hadronic distortion is treated in the following section.

In order to simplify  notation we drop the index on $\bmath{q}_1$ and
put $\bmath{q}_1=  \bmath{q}= (\bmath{q}_{\bot},{q}_{\|})$. 
Then, observe that expression (\ref{study}) can be written as
\begin{equation}
		{\cal M}_C^{(B)}( \bmath{q})= \frac{2Z\alpha}
	   {\bmath{q}_{\bot}^2+q_{\|}^2}\ 
	   \bmath{g}\cdot\bmath{q} =
	   \frac{-1}{2\pi i}\int \rd^3r e^{-i\mathbf{q}\cdot\mathbf{r}}
	     \bmath{g}\cdot\bmath{\nabla}V_C(r) , \label{study-int}
\end{equation}
where $V_C(\bmath{r})$ is the Coulomb-point-nucleus potential
\begin{equation}
	V_C(\bmath{r})=-\frac{Z\alpha}{r}, \label{Point-Cpot}
\end{equation}
and where $\bmath{g}\cdot\bmath{r}=\bmath{g}\cdot\bmath{b}$ with 
$\bmath{r}_{\bot}=\bmath{b}$. 

The Coulomb distortion along the pion trajectory is in the
Glauber model taken into consideration by replacing 
Eq.(\ref{study-int}) by 
\begin{equation}
		{\cal M}_C( \bmath{q})= 
	   \frac{-1}{2\pi i}\int \rd^3r e^{-i\mathbf{q}\cdot\mathbf{r}}
	     \bmath{g}\cdot\bmath{\nabla}V_C(r)\
	      e^{i\chi_C(\mathbf{b})}, \label{Coul-def-gen}
\end{equation}
where $\chi_C(\bmath{b})$ is the Coulomb phase function,
\begin{equation}
  \chi_C(\mathbf{b})=  
  \frac{-1}{v }
    \int_{-\infty}^{\infty}\rd z\,V_C(\mathbf{b},z) \ .
 \label{Coul-phase-fcn}
\end{equation} 
This expression for the amplitude is equally valid for 
extended nuclear-charge
distributions provided the Coulomb potential is
 evaluated with the proper charge
distribution \cite{RJG}.

We first investigate the case of {\it point-like} nuclear charge. 
The Coulomb potential is then as in Eq.(\ref{Point-Cpot})
and the corresponding Glauber expression for the Coulomb 
phase factor
\begin{equation}
	e^{i\chi_C(b)}=\left( \frac{2a}{b}\right)^{i\eta}
\end{equation}
where $a$ is the cut-off radius in the Coulomb potential. For
$\pi^{-}$-nucleus scattering
\begin{equation}
	\eta =2 Z\alpha/v  .\label{etadef}
\end{equation}
The velocity $v$ can in the following safely be put to unity. Thus, 
Eq.(\ref{study-int}) 
recast to include Coulomb scattering becomes
\begin{equation}
{\cal M}_C = \frac{-Z\alpha}{2\pi i}\int \rd^3r e^{-i\mathbf{q}\cdot\mathbf{r}}
	 \      \frac{\bmath{g}\cdot\bmath{r}}{r^3}
	       \left( \frac{2a}{b}\right)^{i\eta} . \label{study-int-2}	
\end{equation}
Integration over the $z$-variable yields a modified Bessel function.
Integration over the angle of the vector $\bmath{q}$ produces 
a factor $\bmath{g}\cdot\bmath{q}_{\bot}$. We
extract this factor and introduce the notation  $F_{C}(\bmath{q})$
for the remaining factor, which is 
an off-shell-Coulomb-scattering amplitude. Hence,
\begin{equation}
{\cal M}_C =  \bmath{g}\cdot\bmath{q}\ F_{C}(\bmath{q}_{\bot},{q}_{\|})  \label{Coul-amp-fact}	
\end{equation}

The Coulomb-scattering amplitude is an integral over impact parameter
\begin{equation}
	F_{C}(\bmath{q})=2Z\alpha /{q}_{\bot}
	     \int_0^{\infty} \rd bJ_1(q_{\bot}b) \{ q_{\|}b K_1(q_{\|}b)\}
   \left( \frac{2a}{b}\right)^{i\eta}. \label{def-Frad}	
\end{equation}
It is convenient to split off the point-Coulomb factor, writing
\begin{equation}
	F_{C}(\bmath{q})= 
	\frac{2Z\alpha (aq)^{i\eta}e^{i\sigma_\eta }}{\bmath{q}^2}\ 
	 h_C(\bmath{q}) \label{FF-with-phase}
\end{equation}
with $\eta$ defined in Eq.(\ref{etadef}) and
\begin{equation}
	\sigma_\eta =2 \arg \Gamma(1-i\eta/2) .
\end{equation}
The extracted phase factors in Eq.(\ref{FF-with-phase}) 
are the same as in  elastic Coulomb
scattering, except that  now 
\begin{equation}
	q=\sqrt{\bmath{q}_{\bot}^2 +  q_{\|}^2} .
\end{equation}
In high-energy-elastic scattering the longitudinal-momentum transfer 
$q_{\|}$ vanishes. In that case $q$  of Eq.(\ref{FF-with-phase})
is interpreted  as ${q}_{\bot}$.

The integration over impact parameter in Eq.(\ref{def-Frad}) leads to 
a hypergeometric function. After some manipulations a simple result 
for the form factor $h_C(\bmath{q})$ emerges
\begin{eqnarray}
	 h_C(\bmath{q})&=& \bmath{q}^2/{q}_{\bot}
	  (aq)^{-i\eta}e^{-i\sigma_\eta }
	     \int_0^{\infty} \rd bJ_1(q_{\bot}b) \{ q_{\|}b K_1(q_{\|}b)\}
   \left( \frac{2a}{b}\right)^{i\eta} \nonumber \\
   &&\nonumber \\
   &=& \Gamma(2-i\eta/2) \Gamma(1+i\eta/2)
	    F(i\eta/2, 1-i\eta/2 ;2; \frac{q_{\bot}^2}{q_{\bot}^2+q_{\|}^2}) .
	      \label{def-FF-coul}
\end{eqnarray}

There are three values of the momentum transfer where the value
of $h_C(\bmath{q})$ is both simple and interesting;
\begin{eqnarray}
	h_C(q_{\bot},q_{\|}=0)&=&  1,\label{FF-arg1} \\
	h_C(q_{\bot}=q_{\|})&=&  
	   \frac{\Gamma(1-i\eta/2)\Gamma(1+i\eta/2)}
	        {\Gamma(1+i\eta/4)\Gamma(\half-i\eta/4)} \sqrt{\pi}, \label{FF-arg2}\\
	h_C(q_{\bot}=0,q_{\|})&=& (1-i\eta/2) \frac{\pi\eta/2}{\sinh(\pi\eta/2)} ;
	  \label{FF-arg3}
\end{eqnarray}
corresponding to $z=1,$ $\half$, and 0.
The value in Eq.(\ref{FF-arg1}) applies to elastic scattering, 
and bremsstrahlung when the transverse momentum transfer is considerably
larger that the longitudinal-minimum-momentum transfer.
The value in Eq.(\ref{FF-arg2}) applies to bremsstrahlung at the
peak where $q_{\bot}=q_{\|}$. The value in Eq.(\ref{FF-arg3}) applies 
to bremsstrahlung in the very forward direction where $q_{\bot}=0$. 
In Table \ref{tab:Num} we give numerical values for three nuclei.
\begin{table}[t]
	\centering
		\begin{tabular}{|c|c|c|c|}
		 \hline
		 Nucleus   & z=1.0 & z=0.5 & z=0.0 \\ \hline
		 C        & 1.0 & 0.998-i0.030 & 0.997-i0.044 \\ \hline
		 Fe       & 1.0  & 0.963 -i0.125 & 0.943 -i0.179 \\ \hline
		Pb         & 1.0 & 0.717 -i 0.272 & 0.588-i0.352 \\ \hline
		\end{tabular}
	\caption{Numerical values of the Coulomb form factor $h_C(z)$
	   with $z=q_{\bot}^2/(q_{\bot}^2+ q_{\|}^2)$.}
	\label{tab:Num}
\end{table}
In the cross-section distribution it is $|h_C(z)|^2$ that enters.

The form factor $h_C(z)$ , which could reduce the 
cross section at the Coulomb peak
by as much as 50 \%, has not always been included. The cross-section 
distributions in \cite{FT3}, e.g., are calculated in the Born approximation.
In order to be valid in the very forward region of 
$q_{\bot}\approx q_{\|}$ those distributions should be multiplied 
by $|h_C(z)|^2$. 
The analysis of the Dubna experiment \cite{Ant} was also done 
without explicitly mentioning this factor.
\begin{figure}[ht]\label{FF-fig}\begin{center}
\scalebox{0.40}{\includegraphics{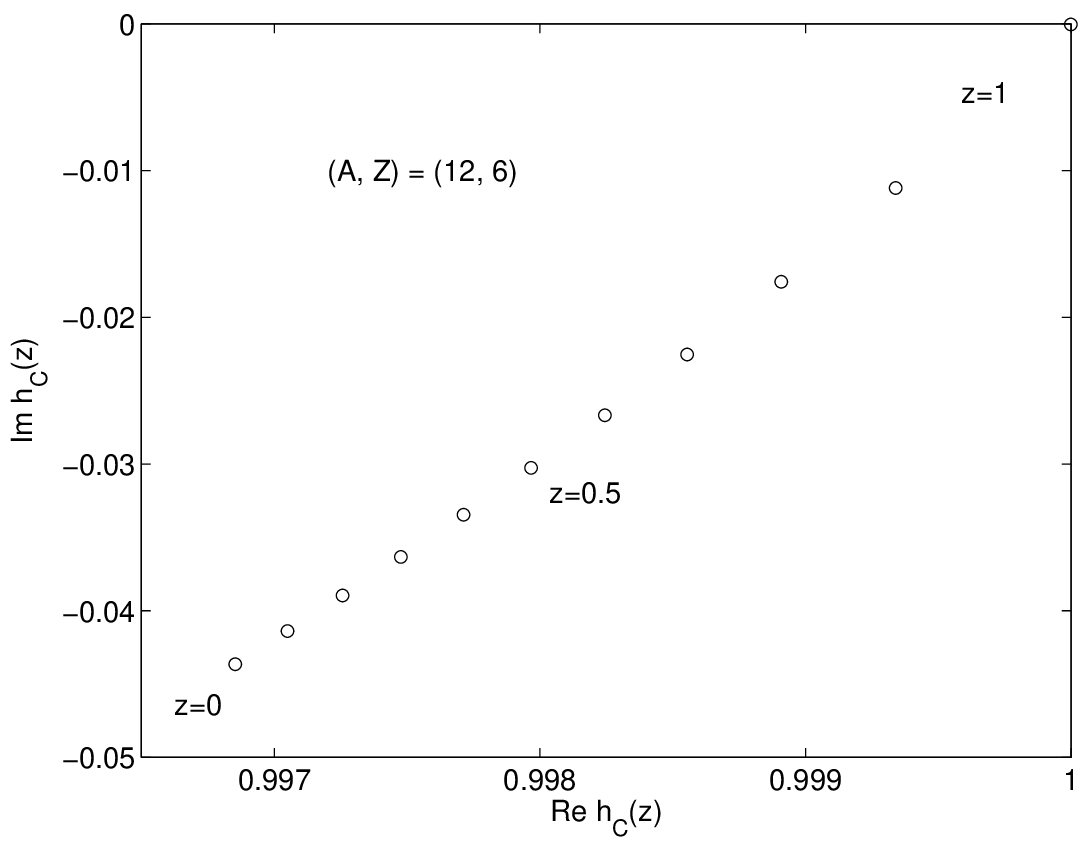} 
\quad\qquad\includegraphics{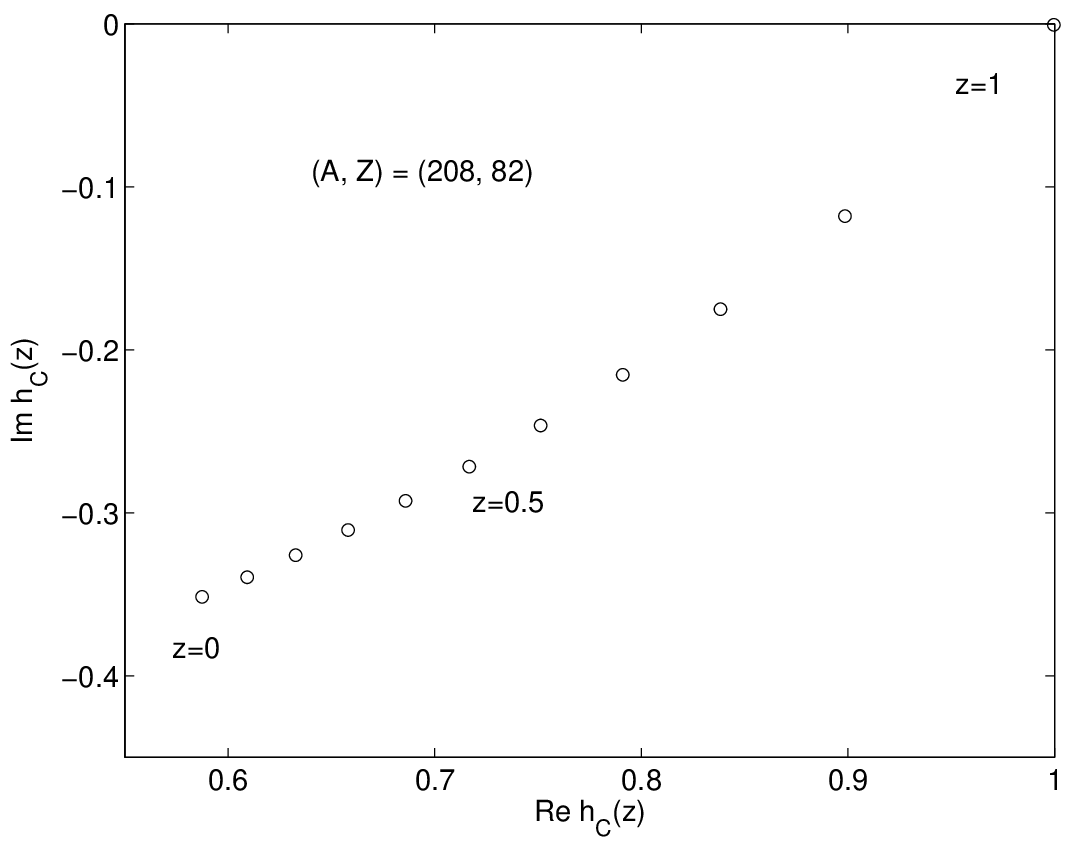}}
\end{center}
\caption{Plots of the Coulomb-form factor \textsf{hc(z)}$=h_C(z)$ 
of Eq.(\ref{def-FF-coul}) for carbon and lead. The 
circles mark points for values of $z$ from 0 to 1.0,
in steps of 0.1. }
\end{figure}

The above expressions  for the 
Coulomb contribution to pionic bremsstrahlung are valid
for point-like nuclear-charge distribution.
The expression for the amplitude ${\cal M_C}$ was given in 
Eq.(\ref{Coul-amp-fact}) and for the form factor 
$F_C$ in Eq.(\ref{def-Frad}). 
This amplitude is summarized by the formula
\begin{equation}
		{\cal M}_C = 
	   \bmath{g}\cdot\bmath{q}\ F_C(\bmath{q}_{\bot},q_{\|})  .\label{M-coul}
\end{equation}
The pion-nucleus-Coloumb-scattering amplitude differs 
slightly from the corresponding elastic amplitude since
off-shell effects have been included through $q_{\|}$.
 
The finite extension of the nuclear-charge distribution 
can also be handled. We replace
the point-Coulomb potential of Eq.(\ref{Point-Cpot}) by the Coulomb
potential $V_C^u(\bmath{r})$, obtained from the extended-charge
distribution. As a result Eq.(\ref{M-coul}) is replaced by
\begin{equation}
		{\cal M}_C = 
	   \bmath{g}\cdot\bmath{q}\ F_C^u(\bmath{q}_{\bot},q_{\|})  ,\label{M-u-coul}
\end{equation}
where $F_C^u(\bmath{q}_{\bot},q_{\|})$ is the  Coulomb-scattering
amplitude of the extended-charge distribution.

The Coulomb-scattering amplitude for an extended-charge distribution cannot be
calculated analytically. Therefore we divide 
the calculation into two steps, writing
\begin{eqnarray}
	F_C^u(\bmath{q}_{\bot},q_{\|}) &=& F_C^p(\bmath{q}_{\bot},q_{\|})
	              +\delta F_C^u(\bmath{q}_{\bot},q_{\|}) \label{F-C-decomp} \\
	     \delta F_C^u(\bmath{q}_{\bot},q_{\|}) &=&
	        F_C^u(\bmath{q}_{\bot},q_{\|}) -F_C^p(\bmath{q}_{\bot},q_{\|}) ,
	        \label{Coul_uniform_F}
\end{eqnarray}
where $F_C^p$ is the point-like-form factor of Eq.(\ref{FF-with-phase}). 
The advantage of this rearrangement
 is that $\delta F_C^u(\bmath{q}_{\bot},q_{\|})$ is a smooth 
function of ${q}_{\bot}$ and $ q_{\|}$ and easily calculated numerically, and 
may in our bremsstrahlung application be evaluated at $ q_{\|}=0$.

From expression (\ref{Coul-def-gen}) for the bremsstrahlung amplitude
we conclude that 
\begin{equation}
	   \bmath{g}\cdot\bmath{q}\ \delta F_C(\bmath{q}_{\bot},q_{\|})
	   = \frac{-1}{2\pi i}\int \rd^3r  
	 e^{-i\mathbf{q}\cdot\mathbf{r}} 
	\left[\bmath{g}\cdot\bmath{\nabla} V_C^u(r)e^{i\chi_C^u(\mathbf{b})}
	  -\bmath{g}\cdot\bmath{\nabla}V_C(r)e^{i\chi_C(\mathbf{b})}\right] ,
	  \label{Def-deltaF}
\end{equation}
where superscript $u$ indicates potential and Coulomb 
phase of the the extended-charge distribution. We assume the nuclear charge 
to vanish outside a radius of $R_u$.

Rearrange   the integrand as follows,
\begin{eqnarray}
	\bmath{g}\cdot\bmath{q}\ \delta F_C^u(\bmath{q}_{\bot},q_{\|})
	&= &\frac{-1}{2\pi i}\int \rd^3r
	 e^{-i\mathbf{q}\cdot\mathbf{r}} 
	\left[ \bmath{g}\cdot\bmath{\nabla}(V_C^u(r)-V_C(r)) 
	   e^{i\chi_C^u(\mathbf{b})}\right.
	\nonumber \\
	 && +\bmath{g}\cdot\bmath{\nabla}V_C(r)
	  \left.
	   ( e^{i\chi_C^u(\mathbf{b})}-e^{i\chi_C(\mathbf{b})})\right] .
	   \label{Coul-corr}
\end{eqnarray}
Then, the first term of the integrand vanishes for $r\geq R_u$, and
since $R_u q_\| \ll 1$ we conclude that the dependence on
$q_\|$ is so weak it can be ignored.
In the second term we integrate over the $z$-variable
and end up with a factor $bq_\| K_1(bq_\|)$. But the
difference between the phase factors vanishes identically
for  $b\geq R_u$, so that everywhere $bq_\| \ll 1$. 
Again it is permissable to take the limit  $q_\|\rightarrow 0$.

The result  of this deliberation is that in 
Eq.(\ref{Def-deltaF}) we can put $q_\|=0$ and get
\begin{eqnarray}
	\delta F_C^u(\bmath{q}_{\bot},q_{\|}) &=&
	   iv\int_0^{R_u} \rd b\ b^2 \frac{J_1(b q_{\bot})}{b q_{\bot}}
	     \ \partial_b\left[ e^{i\chi_C^u(\mathbf{b})}
	  -e^{i\chi_C(\mathbf{b})}\right] \\
	  &=& -iv  \int_0^{R_u} \rd b \ b J_0(q_\bot b) 
	   \left[ e^{i\chi_C^u(b)}
	  -e^{i\chi_C(b)}\right]. \label{Coul_diff_ff}
\end{eqnarray}
The first version is the one best suited for numerical
evaluation. The second version shows explicitely that 
$\delta F_C^u$ is the difference between the Coulomb amplitudes
for extended- and point-charge distributions.

In Fig.~2 we compare the three functions $F_C^u$,  $F_C^p$, and $\delta F_C^u$.
We have chosen $q_\|=1.0$ MeV/$c$, a longitudinal-momentum transfer 
typical for the COMPASS  experiment \cite{COMP}. This 
number is so small that the position of the peak, at $q_\bot =q_\|$, 
cannot be seen in the figure where the curves plotted start
at $q_\bot=10$ MeV/$c$. As is evident, the point-like form factor 
is a good approximation to the uniform form factor up to about 
$q_\bot^2=0.001$ (GeV/$c)^2$.
\begin{figure}[ht]\label{CC-fig}\begin{center}
\scalebox{0.40}{\includegraphics{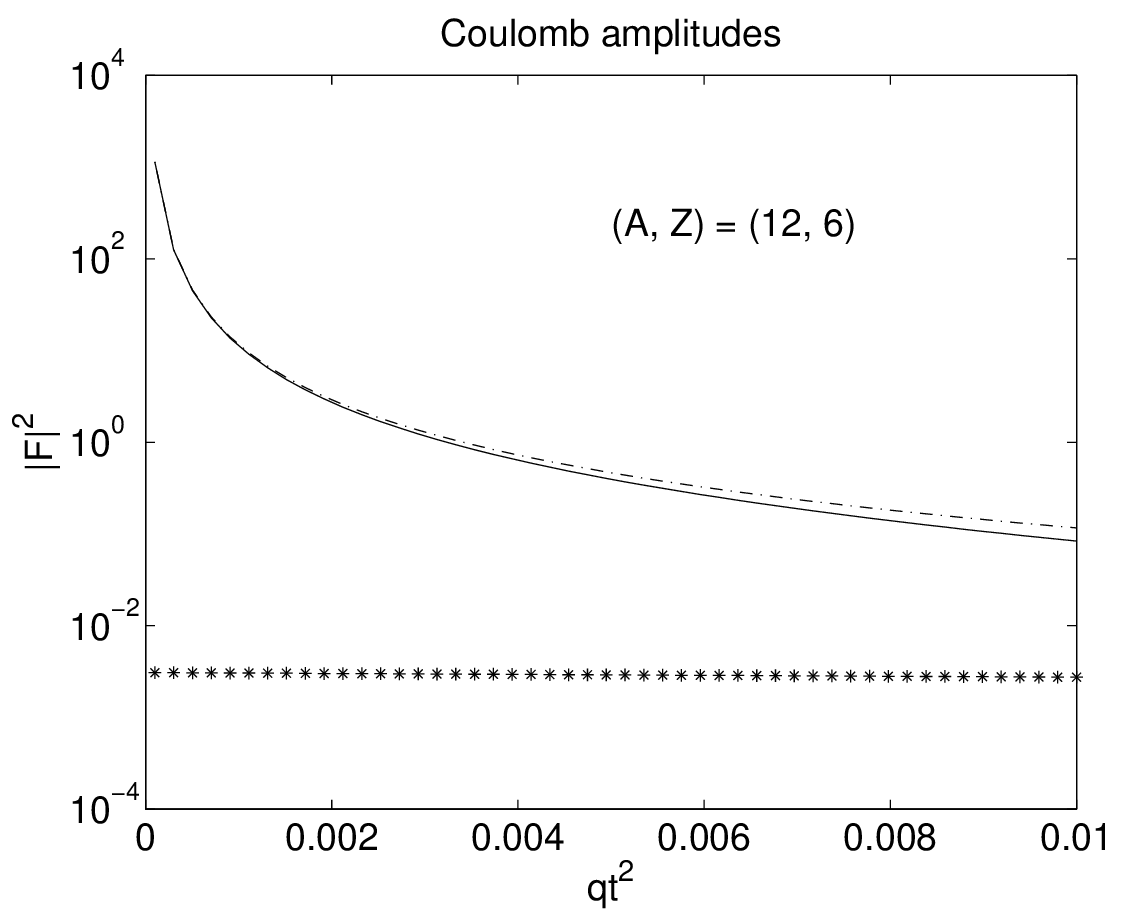} 
\qquad\includegraphics{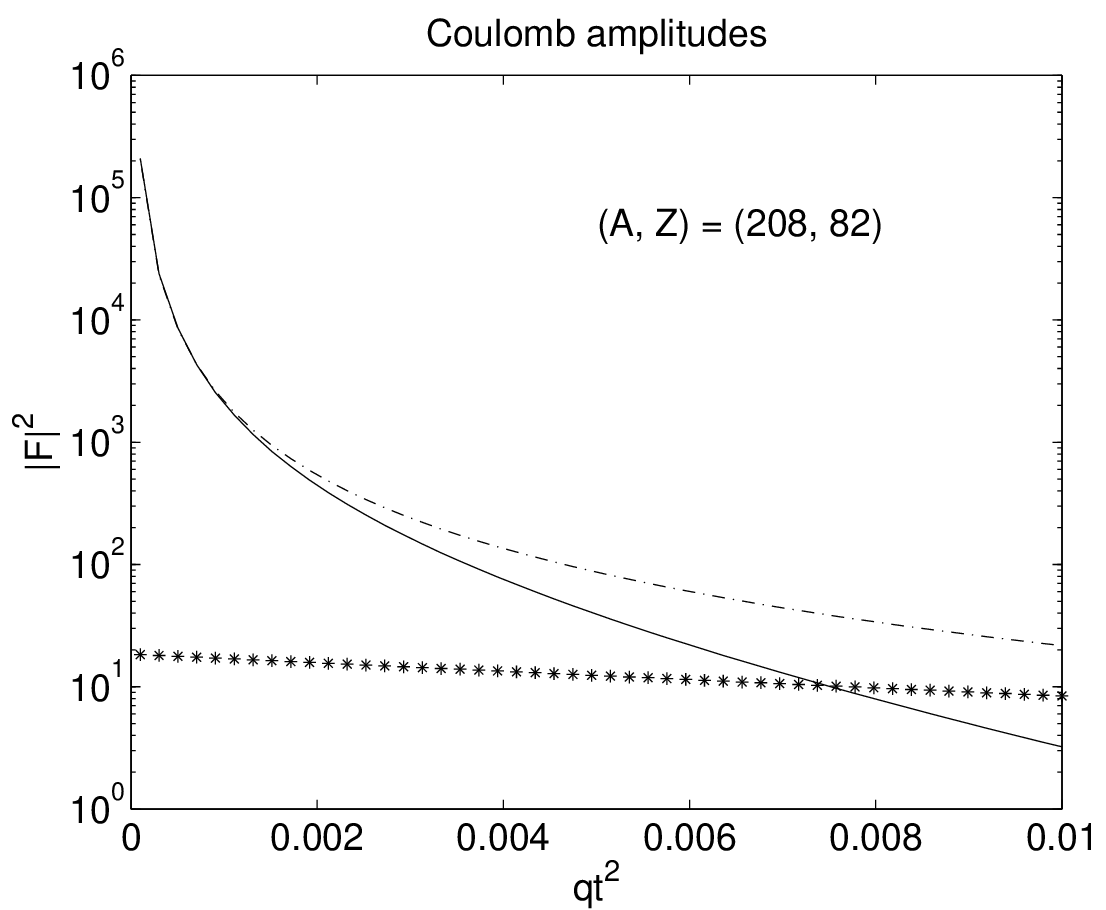}}
\end{center}
\caption{Plots of the squared Coulomb form factors 
for $q_\|=1.0$ MeV/$c$. The dashed curve corresponds the point-like form
factor $F_C^p(\bmath{q}_{\bot},q_{\|})$ of Eq.(\ref{FF-with-phase}),
the starred curve to the difference form factor $\delta F_C^u(\bmath{q}_{\bot},q_{\|})$
of Eq.(\ref{Coul_diff_ff}), and the solid line to their sum, 
the Coulomb form factor $F_C^u(\bmath{q}_{\bot},q_{\|})$ of Eq.(\ref{Coul_uniform_F}). 
The unit for \textsf{qt}$^2=\bmath{q}_{\bot}^2$ is (GeV/$c)^2$.}
\end{figure}

We end this section by remarking that the Coulomb form factor discussed
above is also encountered in  ordinary Coulomb production
\[
a +A\rightarrow a^{\star} + A  ,
\]
where the longitudinal-momentum transfer is defined
as
\begin{equation}
	q_{\|}= (m_{a^{\star}}^2-m_a^2)/(2k),
\end{equation}
with $k$  the momentum of the incident particle $a$.  
In Refs \cite{Mull,Bemp}, and similar applications,
the form factor is calculated numerically, but it is of course
valuable to have an analytic expression for the point-like case.
 In many applications the dependence on $q_{\|}$ 
 is very important, in contrast to the
high-energy bremsstrahlung case discussed here. 
%
%
\newpage
\section{The nuclear amplitude}

It is important to have the correct phase between Coulomb and
nuclear contributions. This point is treated in detail in Ref.\cite{FT1}. 
Suppose the incident pion radiates a photon before scattering. 
Then the nuclear scattering is of course the compounded
amplitude of Coulomb and nuclear scatterings. 
In Eq.(\ref{SimpleII-amp}) the contributions from radiation
from external legs are summarized by the Born approximation
to the pion structure functions, $A=1$ and $B=0$,
changing $\bmath{g}$ into $\bmath{g}_0$  in 
Eq.(\ref{study}).
We want to
extend this contribution by adding the nuclear scattering.
To this end we simply replace the Coulomb potential $V_C(\bmath{r})$
by the sum   $V_C(\bmath{r})+V_N(\bmath{r})$. As for the nuclear
potential we assume the hadronic interaction between 
pion and nucleus to be the same for incident and emerging
pions, even though their energies may be quite different. This
is equivalent to saying that, within the Glauber model,
we assume the pion-nucleon-cross section to be energy 
independent. This
assumption can of course be relaxed.

The compounded nuclear and Coulomb  amplitude corresponding to
Eq.(\ref{study-int}) thus reads
\begin{equation}
  {\mathcal M}(\bmath{q})=  
  \frac{-i}{2\pi }
    \int\rd^3x\, e^{-i\mathbf{q}\cdot\mathbf{x}}\ 
    \mathbf{g}_0\cdot\mathbf{\nabla}
     \left(V_C(\bmath{r})+V_N(\bmath{r})\right)
    e^{i(\chi_C(\mathbf{b})+\chi_N(\mathbf{b}))} .
 \label{FGl_C+N}
\end{equation}
The momentum transfer $\mathbf{q}$ is  three-dimensional and the 
distortion includes both Coulomb and hadronic distortion. The 
relation between potentials and phase-shift functions is 
defined in Eq.(\ref{Coul-phase-fcn}). 

The integrand of Eq.(\ref{FGl_C+N}) can be rearranged to read
\begin{equation}
  \mathbf{g}_0\cdot\mathbf{\nabla}V_C(\bmath{x})e^{i\chi_C(\mathbf{b})}
  +\mathbf{g}_0\cdot\mathbf{\nabla}V_N(\bmath{x})        
  e^{i(\chi_C(\mathbf{b})+\chi_N(\mathbf{b}))}
  -  \mathbf{g}_0\cdot\mathbf{\nabla}V_C(\bmath{x})e^{i\chi_C(\mathbf{b})}
   (1-e^{i\chi_N(\mathbf{b})}) .
 \label{integrand}
\end{equation}
The three terms are quite different in nature. The integrand of the 
first term extends
 over all of space, since the Coulomb potential does. The integrand of 
 the second term 
 is non-zero only inside the nucleus, since only there is the 
 nuclear potential non-vanishing. Also the integrand of the
 third term  vanishes
 outside the nucleus, since the factor $(1-e^{i\chi_N(\mathbf{b})})$
 there does.
 
 The first term of Eq.(\ref{integrand}) describes  the Coulomb 
 contribution, for a general charge distribution. In the second and third 
 terms
 we can neglect the functional dependence on the longitudinal 
 momentum transfer, since $q_{\|}$ is fixed and so small that 
 $R_u q_{\|}\ll  1$ for all nuclei. The nuclear contribution to the 
 bremsstrahlung amplitude, i.e.~the second and third terms of 
 Eq.(\ref{integrand}), can be written as 
 \begin{eqnarray}
  {\mathcal M}_{N}(\bmath{q})&=&  \mathbf{g}_0\cdot\mathbf{q}\ 
                   F_N(\mathbf{q}_\bot) ,\\
   F_N(\mathbf{q}_\bot) &=& 
  \frac{iv}{2\pi }
    \int\rd^2 b\, e^{-i\mathbf{q}_\bot\cdot\mathbf{b}}
    e^{i\chi_C(\mathbf{b})} \left[1- e^{i\chi_N(\mathbf{b})}\right] .
 \label{F_N}
\end{eqnarray}
The  factor $F_N(\mathbf{q}_\bot)$ is simply the elastic pion-nucleus 
scattering 
amplitude divided by the energy. It is energy independent since
we assumed energy independent pion-nucleus interactions.
\begin{figure}[ht]\label{CN-fig}\begin{center}
\scalebox{0.40}{\includegraphics{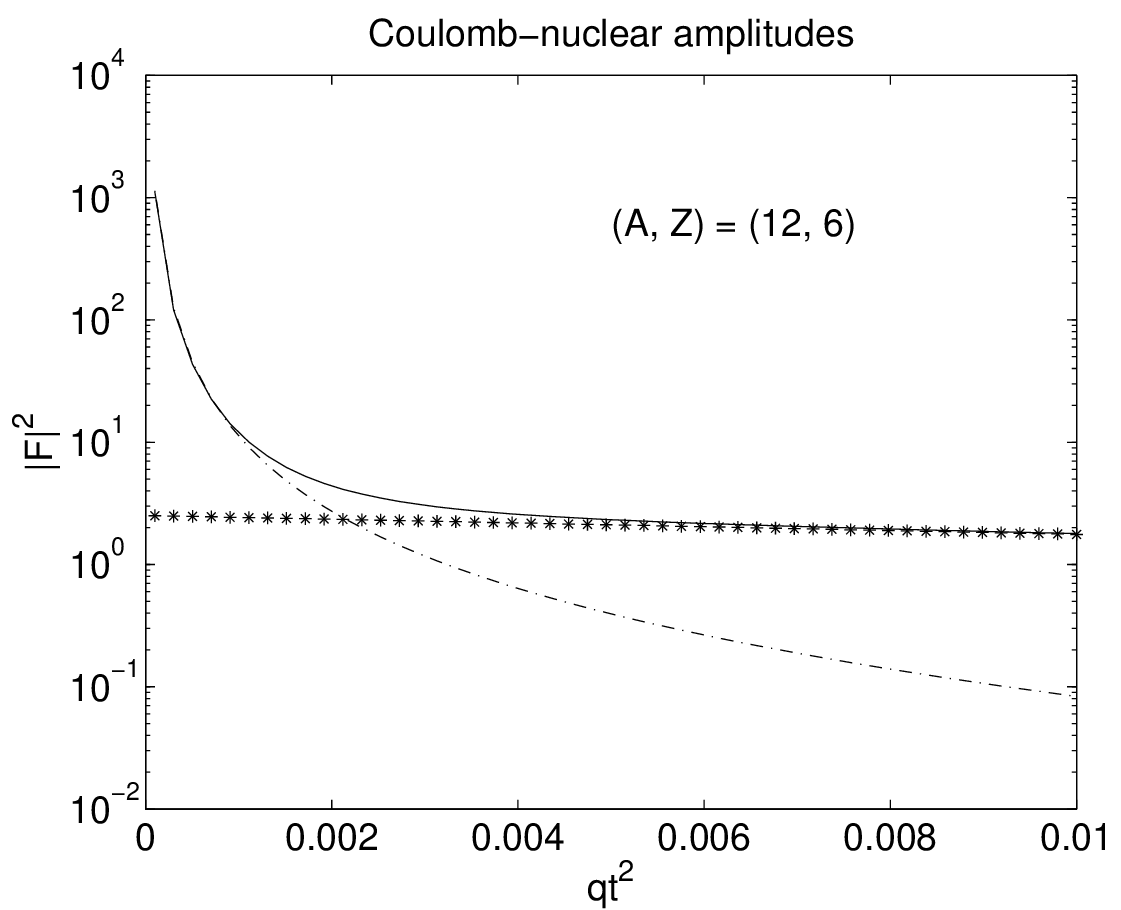} 
\quad\qquad\includegraphics{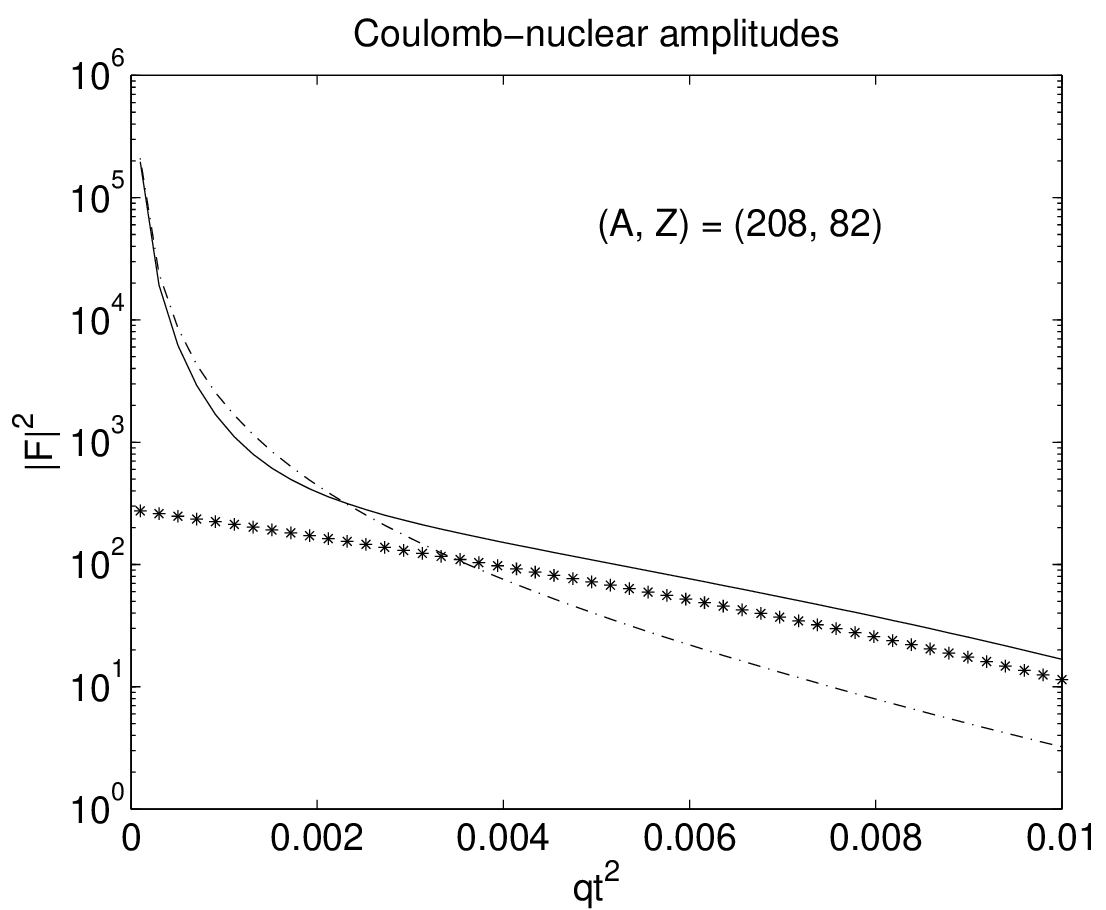}}
\end{center}
\caption{Plots of the squared form factors. The dashed curve corresponds the 
uniform-Coulomb-form factor $F_C^u(\bmath{q})$ of Eq.(\ref{Coul_uniform_F}),
the starred curve to the nuclear-form factor $ F_N(\bmath{q})$
of Eq.(\ref{F_N}), and the solid line to their sum. The unit for 
\textsf{qt}$^2=\bmath{q}_{\bot}^2$ is (GeV/$c)^2$. }
\end{figure}

In Fig.~3 we have plotted what essentially amounts to 
elastic pion-nucleus cross-section distributions. The dashed lines represent
Coulomb scattering, the starred lines nuclear scattering, and the solid
lines their sum. In all terms we have neglected the longitudinal-momentum
transfer, being so incredibly small on the scale of momenta
 plotted. For transverse momentum transfers
$q_\bot^2\geq 0.002$ (GeV/$c)^2$ the hadronic contribution dominates 
the Coulomb contribution.
%
%
%
\newpage
\section{The polarizability amplitude}
Now, we  have the complete amplitude for point-like pions. But
the aim is to incorporate the pion polarizabilities, which are
represented by the vector $\mathbf{g}-\mathbf{g}_0$ in 
the one-photon-exchange matrix element of Eq.(\ref{study}).
Including Coulomb and hadronic distortions gives, instead of 
Eq.(\ref{Coul-def-gen}), the polarizability amplitude
\begin{equation}
		{\cal M}_P( \bmath{q})= 
	   \frac{-1}{2\pi i}\int \rd^3r e^{-i\mathbf{q}\cdot\mathbf{r}}
	     (\bmath{g}-\bmath{g}_0)\cdot\bmath{\nabla}V_C(r)\
	       e^{i(\chi_C(\mathbf{b})+\chi_N(\mathbf{b}))}. \label{M-pol}
\end{equation}
Following our welltrodden path we rewrite the distortion factor as
\begin{equation}
	e^{i(\chi_C(\mathbf{b})+\chi_N(\mathbf{b}))} =
	  e^{i\chi_C(\mathbf{b})} -e^{i\chi_C(\mathbf{b})}(1-e^{i\chi_N(\mathbf{b})}) .
\end{equation}
The first term in this decomposition yields upon integration the Coulomb scattering
amplitude $F_C(\mathbf{q})$. The second term vanishes for impact parameters
 \begin{figure}[b]\label{CP-fig}\begin{center}
\scalebox{0.40}{\includegraphics{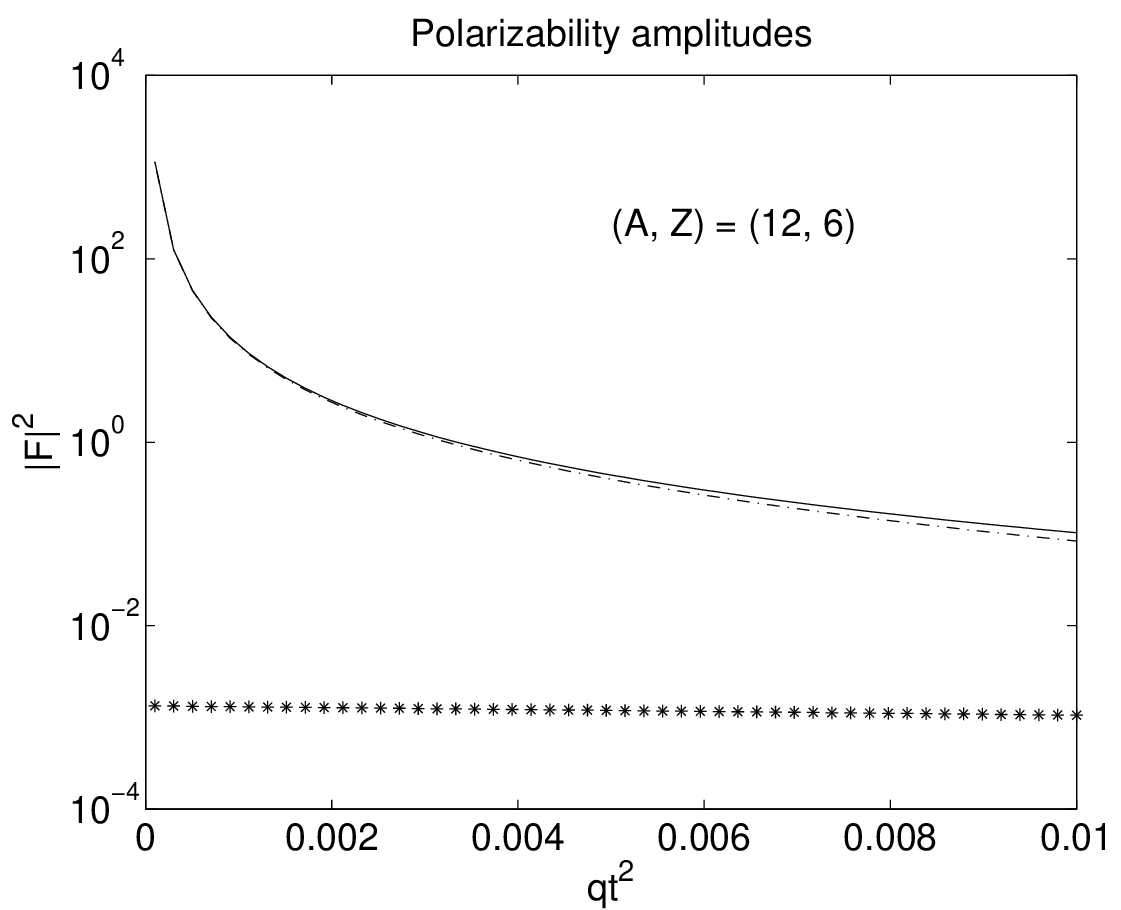} 
\quad\qquad\includegraphics{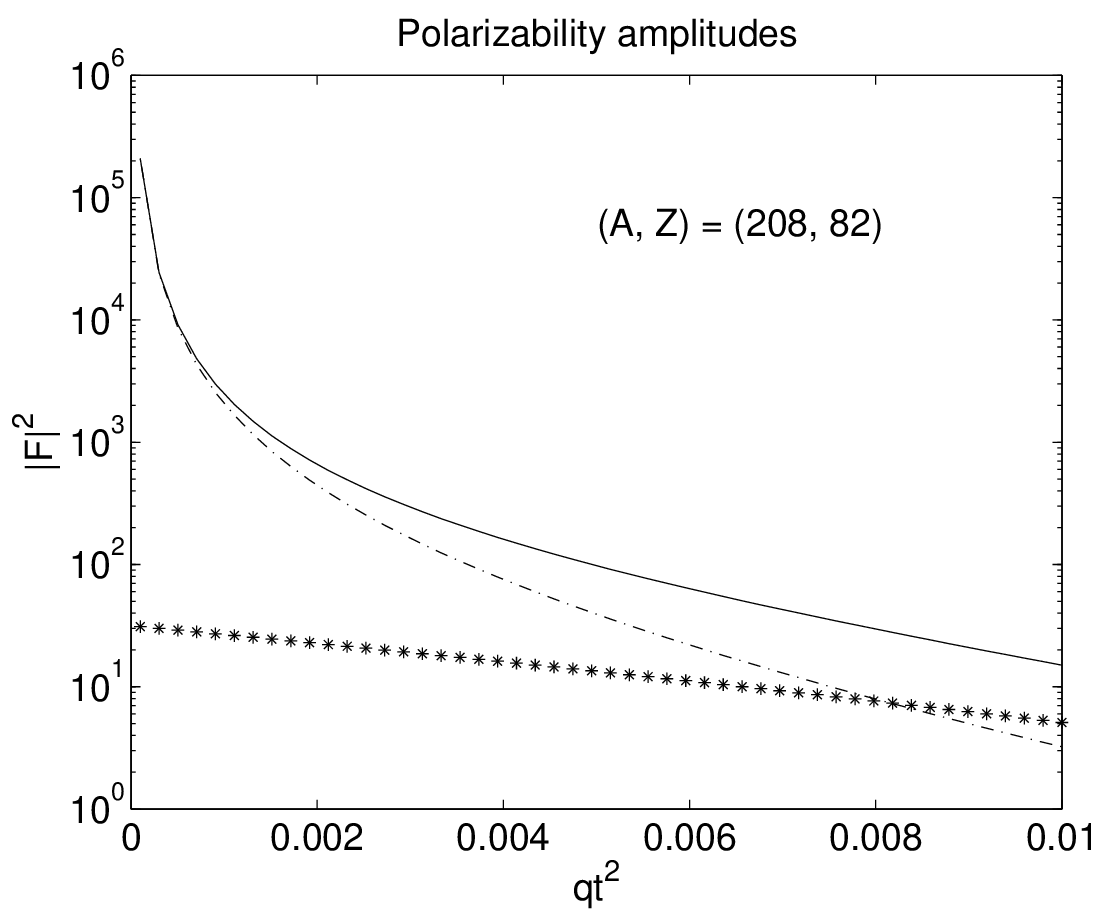}}
\end{center}
\caption{Plots of the squared form factors. The dashed curve corresponds the 
uniform-Coulomb-form factor $F_C^u(\bmath{q})$ of Eq.(\ref{Coul_uniform_F}),
the starred curve to the hadronic-distortion-form factor $ \delta F_P(\bmath{q}_{\bot})$
of Eq.(\ref{deltaF_P}), and the solid line to their sum. 
The unit for \textsf{qt}$^2=\bmath{q}_{\bot}^2$ is (GeV/$c)^2$. }
\end{figure}
$b\geq R_u$, and leads to a smooth term, as discussed above, where we
can take the limit $q_\|\rightarrow 0$. Our result for the 
polarizability contribution to the bremsstrahlung amplitude is
therefore
\begin{eqnarray}
		{\cal M}_P( \bmath{q}) &=&
		(\bmath{g}-\bmath{g}_0)\cdot\bmath{q} F_P(\mathbf{q}),\\
		F_P(\mathbf{q}) &=& F_C^u(\mathbf{q})+  \delta F_P(\mathbf{q}_\bot)	 \label{F_P} ,
\end{eqnarray}
with the hadronic contribution 
\begin{equation}
\delta F_P(\mathbf{q}_\bot)	  =
	   iv\int_0^{R_u} \rd b\ b^2 \frac{J_1(b q_{1\bot})}{b q_{1\bot}}
	      \left[\partial_b  e^{i\chi_C^u(\mathbf{b})} \right]
	 \left(1- e^{i\chi_N(\mathbf{b})}\right) . \label{deltaF_P}
\end{equation}
This amplitude is a smooth function of $\mathbf{q}_\bot$.
The fact that the hadronic distortion effects are quite different for the Born
 and  the polarizability amplitudes was raised already in Ref.\cite{FT1}. 

In Fig.~4 we graph the squared polarization-form factors. The dashed curve 
represents the Coulomb-scattering-form factor,  
 the starred curve the hadronic-distortion-form factor,
 and the solid curve their sum. We see that  hadronic
 effects are much weaker for the Compton polarizability amplitude
 than for the Compton Born amplitude, a conclusion that should be
 evident from expression (\ref{deltaF_P}), being proportional
 to the electromagnetic coupling as it is. Another conclusion
 that can be drawn from Fig.~4 is that 
 the strength of the polarizability amplitude  compared with 
 that of the Born amplitude  diminishes 
 when the transverse momentum transfer moves into the 
 region $q_\bot^2\geq 0.002$ (GeV/$c)^2$.

%
%
%
\newpage
\section{Bremsstrahlung cross-section distribution}
The complete pion-nucleus bremsstrahlung amplitude has 
a simple structure
\begin{equation}
  {\mathcal M}= \bmath{g}_0\cdot \bmath{q}_{1\bot}\ F_C(\bmath{q}_1)
  +(\bmath{g}-\bmath{g}_0) \cdot \bmath{q}_{1\bot}\ F_P(\bmath{q}_1)
    +  \bmath{g}_0\cdot \bmath{q}_{1\bot}\ F_N(\bmath{q}_{1\bot}) , 
    \label{Sum-amp}
\end{equation}
where $F_C$ is the off-shell pion-nucleus Coulomb scattering
amplitude, $F_N$ the on-shell pion-nucleus hadronic
scattering amplitude, and $F_P$ a mixed amplitude appropriate
for the polarizability contribution. Moreover, $F_P$ is essentially 
equal to $F_C$. The fact that the Coulomb amplitude is off-shell
is only important in the region of the Coulomb peak, where $\bmath{q}_{1\bot}$
is of a size similar to the constant ${q}_{1\|}=q_{min}$.

It is straightforward to calculate the cross-section distribution 
from Eq.(\ref{Sum-amp}). However, in practice the polarizability
contributions are small, and in the expressions below it
is often sufficient to  keep the corresponding linear terms.
After summation over the polarization directions of the final state photon
we get for the cross-section distribution of Eq.(\ref{Cross-sect-distr})
\begin{equation}
\frac{\rd \sigma}{\rd^2q_{1\bot}  \rd^2q_{2\bot} \rd x}
  = \frac{\alpha\ \mathbf{q}_{1\bot}^2 }{\pi^2m_{\pi}^4}
   \Bigg( \frac{1-x}{x^3} \Bigg) 
   \left( \frac{x^2 m_{\pi}^2}{\mathbf{q}_{2\bot}^2+x^2 m_{\pi}^2}\right)^2 
  \Bigg({\mathcal K}_1 + {\mathcal K}_2 +  {\mathcal K}_3 \Bigg)  ,
 \label{Cross-sect}
 \end{equation}
where ${\mathcal K}_1$ is the Coulomb-nuclear contribution for point-like pions
\begin{eqnarray}
{\mathcal K}_1 &= & \left| F_C(\bmath{q}_1) + F_N(\bmath{q}_1) \right|^2
    \left( 1 - \mu^2\frac{4 x^2 m_{\pi}^2\mathbf{q}_{2\bot}^2}
                          {(x^2 m_{\pi}^2 +\mathbf{q}_{2\bot}^2)^2} \right),
 \label{Coul-nucl-contr}
 \end{eqnarray}
  ${\mathcal K}_2$  the  contributions linear in the pion-polarizability
  functions
\begin{eqnarray}
{\mathcal K}_2 &= & 2\Re 
  \Bigg[\left( F_C^{\star}(\bmath{q}_1)+  F_N^{\star}(\bmath{q}_1)\right)
    F_P(\bmath{q}_1)\Bigg]
    \Bigg[C(x,\mathbf{q}_{2\bot}^2)
    \left( 1 - \mu^2\frac{4 x^2 m_{\pi}^2\mathbf{q}_{2\bot}^2}
                          {(x^2 m_{\pi}^2 +\mathbf{q}_{2\bot}^2)^2} \right)
                             \nonumber \\   
    && \qquad +
        B(x,\mathbf{q}_{2\bot}^2)
    \left( 1 - \mu^2\frac{2 \mathbf{q}_{2\bot}^2}
                          {x^2 m_{\pi}^2 +\mathbf{q}_{2\bot}^2} \right) \Bigg] ,
 \label{Lin-pol-contr}
 \end{eqnarray}
and finally, ${\mathcal K}_3$  the  contributions quadratic
 in the pion-polarizability  functions
\begin{eqnarray}
{\mathcal K}_3 &= & \left| F_P(\bmath{q}_1)\right|^2
   \Bigg[ \left| C(x,\mathbf{q}_{2\bot}^2)\right|^2
    \left( 1 - \mu^2\frac{4 x^2 m_{\pi}^2\mathbf{q}_{2\bot}^2}
                          {(x^2 m_{\pi}^2 +\mathbf{q}_{2\bot}^2)^2} \right)
                + \left|B(x,\mathbf{q}_{2\bot}^2)\right|^2           \\
      && \qquad+2\Re\left(  C(x,\mathbf{q}_{2\bot}^2)
        B^{\star}(x,\mathbf{q}_{2\bot}^2)\right)
    \left( 1 - \mu^2\frac{2 \mathbf{q}_{2\bot}^2}
                          {x^2 m_{\pi}^2 +\mathbf{q}_{2\bot}^2} \right) \Bigg],
 \label{Qua-pol-contr}
 \end{eqnarray}
 where, in order to shorten the expressions, we have introduced
\begin{equation}
	A(x,\mathbf{q}_{2\bot}^2)=1+C(x,\mathbf{q}_{2\bot}^2) .
\end{equation}
 The parameter $\mu$ is defined as
  $\mu=\hat{\mathbf{q}}_{1\bot}\cdot\hat{\mathbf{q}}_{2\bot}$. 
  On the right hand sides of the above formulae we may in most 
  applications replace $\mu^2$ 
  by its average $\half$.
  
 The dominant contribution to the cross section Eq.(\ref{Cross-sect})
 comes from the point-like-pion approximation, i.e.\ the
 contribution proportional to ${\mathcal K}_1$ of 
 Eq.(\ref{Coul-nucl-contr}). The polarizbility contributions are 
 contained in ${\mathcal K}_2$ and ${\mathcal K}_3$. Experiments
 are aimed at measuring the contribution proportional to ${\mathcal K}_2$,
 which is linear in the polarizabilities. The relative nuclear-form factor 
 between the ${\mathcal K}_2$ and ${\mathcal K}_1$ contributions is
\begin{equation}
 R(\bmath{q}_1)= \frac{ F_P(\bmath{q}_1)}{F_C(\bmath{q}_1)+  F_N(\bmath{q}_1)}
 ,
 \label{Pol-ratio}
 \end{equation}
with the polarizability-form factor $F_P(\bmath{q}_1)$ as defined 
in Eq.(\ref{F_P}), and with $\bmath{q}_1$ the momentum transfer 
to the nucleus.  
When hadronic interactions of the pions are neglected, the ratio 
$R(\bmath{q}_1)$  becomes one.  In Fig.~5 we plot this 
ratio  as a function of $\bmath{q}_{1\bot}^2$ in the interval 
$0\leq \bmath{q}_{1\bot}^2 \leq 0.002$ (GeV/$c$)$^2$.
\begin{figure}[h]\label{Ratio-fig}\begin{center}
\scalebox{0.55}{\includegraphics{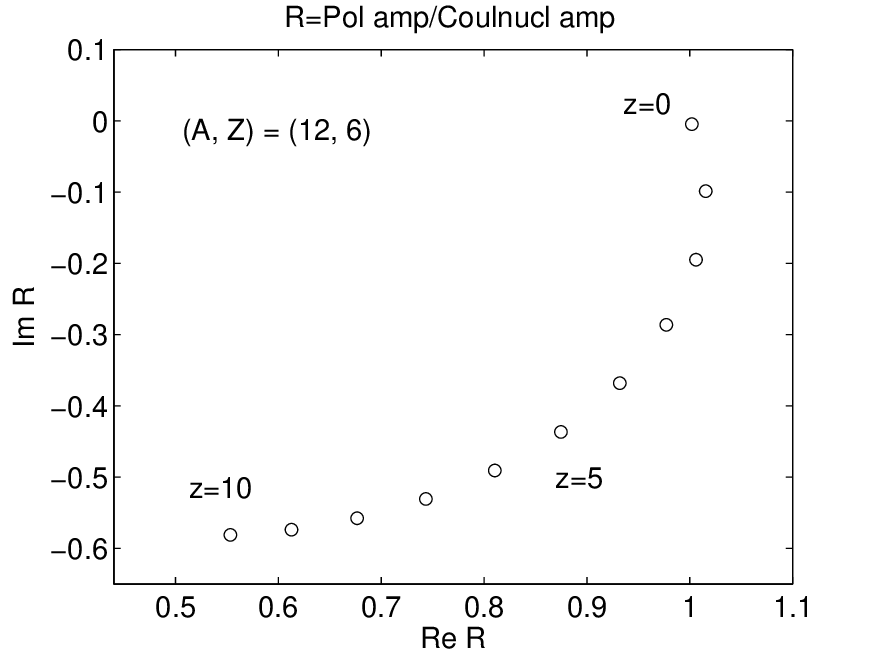} 
\quad\qquad\includegraphics{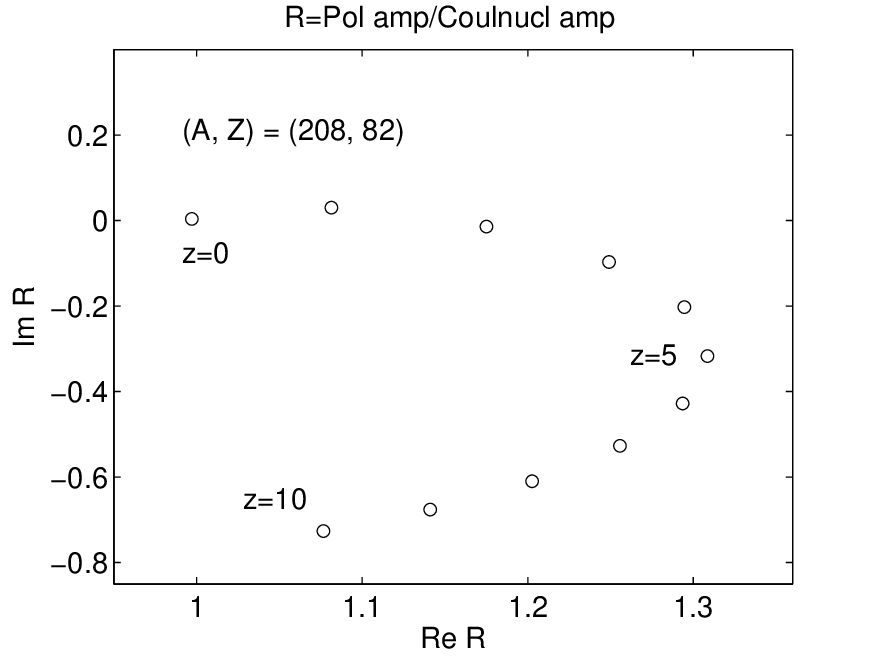}}
\end{center}
\caption{Plots of the ratio  $R(\bmath{q}_1)$ 
of the form factors for the linear polarizability cross-section contribution 
and the point-like-pion cross-section contribution, of Eq.(\ref{Pol-ratio}).
 The circles 
represent the ratios for  $\bmath{q}_{1\bot}^2$ in the interval 
$0\leq \bmath{q}_{1\bot}^2 \leq 0.002$ (GeV/$c$)$^2$ in steps of 
0.0002 (GeV/$c$)$^2$. The value of $R(\bmath{q}_1)$ is one at 
$\bmath{q}_{1\bot}^2=0$. The unit for $\bmath{q}_{\bot}^2$ is (GeV/$c)^2$.}
\end{figure}

In Fig.~5 we have limited the region of $\bmath{q}_{1\bot}^2$, 
since we know 
from the graphs of the previous sections that 
for $\bmath{q}_{1\bot}^2\geq 0.002$ (GeV/$c$)$^2$ 
 the contributions
from the hadronic interactions of the pions play a dominant 
role. Of course, our model is valid also in this case,
but experimenters prefer to stay in the region where 
the description is simple,
meaning $R(\bmath{q}_1)=1$. From Fig.~5 we conclude that if this 
is desired we must further restrict
the momentum transfer to the nucleus. For Compton masses in the 
threshold region, sufficiently below the $\rho$-meson mass, 
the polarizability functions $C(x,\mathbf{q}_{2\bot}^2)$ and 
$B(x,\mathbf{q}_{2\bot}^2)$ of Eq.(\ref{Lin-pol-contr}) are
real-valued functions. Therefore, in the threshold region only the 
real part of $R(\bmath{q}_1)$ matters and   
limiting  ourselves to  
$\bmath{q}_{1\bot}^2 \leq 0.001$ (GeV/$c$)$^2$, it is  reasonable
to set $R(\bmath{q}_1)\approx 1$. In the general case,
however, the more detailed model developed here must be applied.
%
\newpage
\section{Summary}

The pion-Compton scattering amplitude is near threshold fixed by 
Born terms involving pion-exchange diagrams (Thompson scattering). 
At higher energies structure dependent terms enter, 
labeled electric and magnetic polarizabilities (Rayleigh scattering).
Those terms can be modelled as $\sigma$-, $\rho$-, and $a_1$-exchange
contributions. 

Pion-nucleus bremsstrahlung is closely related to pion-Compton 
scattering. At small momentum transfers to the nucleus 
the bremsstrahlung reaction is dominated by single-photon exchange
between the pion and the nucleus.
As a consequence, the bremsstrahlung amplitude becomes proportional to 
the pion-Compton scattering amplitude, the initial photon of the 
Compton scattering being the virtual photon the pion is exchanging
 with the nucleus.

For heavy nuclei multiple-photon exchange becomes important. 
But its sole effect is to introduce the well-known Coulomb 
phase factor. In  the bremsstrahlung reaction the phase 
is slightly different from the one in elastic Coulomb scattering,
since in bremsstrahlung there is a fixed longitudinal momentum
transfer to the nucleus, $q_{min}$. A second effect produced by 
the longitudinal momentum transfer is the appearance 
of a new form factor. An analytic form for this form factor
is given, for the first time. It is important for
heavy nuclei when the transverse momentum transfer to the nucleus
is similar in magnitude to $q_{min}$.

However, pionic bremsstrahlung can also be accompanied by 
pion-nucleus hadronic scattering. The importance of this contribution
increases as the transverse momentum transfer increases, exactly 
as in elastic scattering. It affects both  Born and
polarizability parts of the Compton amplitudes. The  Born term
becomes, essentially, multiplied by the sum of elastic 
Coulomb and hadronic pion-nucleus scattering amplitudes. 
For the polarizability terms there is a corresponding sum, 
but whereas the Coulomb amplitude is the same, the hadronic one
is different and substantially weaker. 

Numerical estimates of the various contributions are presented.
The outcome is that
if one is interested in extracting polarizability contributions,
it is advantageous to restrict oneself to momentum transfers 
$q_{\bot}^2\leq 0.001$ (GeV/$c$)$^2$, since there the ratio
between polarizability and Born contributions remains essentially 
the same as in free pion Compton scattering. Increasing
the momentum transfer means increasing the importance of 
hadronic scattering. The ratio then changes in an important way, 
and varies with momentum transfer. Pushing into the momentum
transfers region $q_{\bot}^2\geq 0.002$ (GeV/$c$)$^2$ we come
into a region where hadronic scattering dominates, and where 
in addition the contribution from the polarizability terms 
diminishes.
%
%
%

\section{acknowlegments}
We are very greatful to Jan Friedrich for pointing out  errors 
in the original expression for the uniform-Coulomb-pase-shift 
functions, Eqs.(51,52). They are correct in 
the equations below. The figures have been recalculated
accordingly. Eqs.(A1,A2) and the figures of the published 
version of this article, Ref.\cite{prcpub}, should be similarly
corrected.
\newpage
\section{Appendix}
In this Appendix we explain how we have calculated the 
Coulomb and nuclear amplitudes. It is then important to remember
that we need the amplitudes only for small momentum transfers,
meaning that the structure of the nuclear surface region
will not be important. Hence, we choose uniform nuclear
charge and matter distributions and with the same radii, 
$R_u=1.1 A^{1/3}$ fm.
More sophisticated calculations are straightforward but also more
time consuming.

We start with the {\it Coulomb amplitude}. The Coulomb phase 
contains a cut-off $a$ that should go to infinity. In this limit
the cut-off enters as a  phase factor common to  
both Coulomb and nuclear amplitudes. The value of $a$ is therefore
immaterial  and we may simply replace $2a$ by $R_u$.
We also put $v=1$.

The Coulomb-phase function for a uniform-charge distribution is
\cite{UnCoul}
\begin{eqnarray}
	\chi_C^u(b)&=& 2Z\alpha \ln (R_u/b),   \qquad b>R_u \nonumber \\
	  &=& 2Z\alpha \Bigg[ \left(1 - b^2/R_u^2\right)^{1/2} 
	   +\frac{1}{3} \left(1 - b^2/R_u^2\right)^{3/2} 
	   \nonumber \\ && \qquad \qquad
	     - \ln\left(1 +\sqrt{1 - b^2/R_u^2}\right) \Bigg]
	      ,   \quad  b<R_u .
	      \label{A1}
\end{eqnarray}
We shall also need the derivatives
\begin{eqnarray}
	b\partial_b \chi_C^u(b)&=& -2Z\alpha ,   \qquad b>R_u \nonumber \\
	  &=& -2Z\alpha \Bigg[ 1-
	        \left(1 - b^2/R_u^2\right)^{3/2}  \Bigg]
	      ,   \qquad b<R_u .
	       \label{A2}
\end{eqnarray}
The Coulomb-phase function for a point-charge distribution is
\begin{equation}
	\chi_C(b)=2Z\alpha \ln (R_u/b) .
\end{equation}

The Coulomb-scattering amplitude $F_C^u(\bmath{q}_1)$ is 
decomposed as in Eq.(\ref{F-C-decomp}). It is written
as a sum of two terms; the point-Coulomb amplitude and a correction term,
$\delta F_C^u(\bmath{q}_1)$. The point amplitude is calculated
exactly. The correction term is the difference between the Coulomb
amplitudes for exteneded and point charges, respectively.
In this term the fixed longitudinal momentum transfer can
be put to zero. The difference is calculated numerically from the
formula
\begin{equation}
	\delta F_C^u( \bmath{q}_{1\bot}, q_{1\|})= i
		\int_0^{R_u}\rd b\  b^2        
	 \frac{J_1(q_{1\bot}b)}{q_{1\bot}b} \ 
	  \partial_b \Bigg[ e^{i\chi_C^u(b)}
	   - e^{i\chi_C(b)}\Bigg] .
\end{equation}
The integral in the last step extends over the nuclear charge 
distribution alone.

Next we look at the {\it nuclear  amplitude} of Eq.(\ref{F_N}).
\begin{equation}
  F_{N}(q_\bot)=  i
    \int_0^{\infty}\rd b\  b J_0(q_{\bot}b) 
    e^{i\chi_C^u(b)} \left[1- e^{i\chi_N(b)}\right] .
 \label{FFNF}
 \end{equation}
 The nuclear phase-shift function is related to the target-thickness
 function $T_A(b)$ by
\begin{equation}
	i\chi_N(b)= -\half \sigma (1-i \alpha )T_A(b) ,
\end{equation}
where $\sigma$ is the pion-nucleon total cross section, and 
$\alpha$ the ratio of real to imaginary part of the forward
elastic pion-nucleon scattering amplitude.
The target-thickness function for a nucleus of uniform density is
\begin{equation}
	T_A(b)=\frac{3A}{2\pi R_u^2}\sqrt{1-b^2/R_u^2} .
\end{equation}
We have chosen numerical values for the hadronic parameters 
appropriate for pions of 190 GeV/$c$; i.e.\ $\sigma=24.1$ mb and
$\alpha=-0.06$ \cite{CARR, BURQ}.

\end{document}